\documentclass[a4paper,11pt]{article}
\pdfoutput=1 

\usepackage{jinstpub} 

\title{The ALICE Muon IDentifier (MID)}

\author[a, 1]{Livia Terlizzi,\note{Corresponding author.}}

\affiliation[a]{INFN and University of Torino, via Pietro Giuria 1, Torino, Italy}

\emailAdd{livia.terlizzi@unito.it}

\abstract{During the LHC Run-I (2010-2013) and Run-II (2015-2018), the selection of interesting events for muon physics in ALICE was performed with a dedicated muon trigger system, composed of 72 single-gap phenolic Resistive Plate Chambers (RPCs) operated with current front-end electronics (ADULT FEE) that provides no amplification for the detector pulses. 
From Run-III starting in 2021 on, in order to fully profit from the increased luminosity of Pb-Pb collisions, the ALICE experiment will run in continuous readout (triggerless) mode and the Muon TRigger system (MTR), therefore, plays the role of a Muon IDentifier (MID). The read-out electronics is being upgraded in order to support continuous readout.
Furthermore, in order to increase the RPC rate capability and to mitigate possible aging effects, it is advantageous to operate the detectors with a lower gain avalanche mode by using higher sensitive FEE. Therefore, we decided to replace the current FEE cards with new ones named Front-End Electronics Rapid
Integrated Circuit (FEERIC) equipped with an amplification stage. 
 Also, an upgrade of the threshold distribution system to the front-end will allow one to tune thresholds at the single front-end card level, while this was previously only possible at the single-RPC level. Finally, since some of the RPCs currently installed in ALICE have accumulated charges significantly comparable to the value that is expected in their lifespan, roughly 25\% of the detectors are planned to be replaced by new ones constructed with new phenolic high-pressurized laminates (HPL). 
A detailed description of the MTR upgrade and of its current status will be presented in this contribution.}


\collaboration[c]{on behalf of the ALICE collaboration}

\proceeding{15$^{\text{th}}$ Workshop on Resistive Plate Chambers and Related Detectors\\
  10-14 February 2020\\
  University of Rome Tor Vergata}

\begin{document}
\maketitle
\flushbottom

\section{Introduction}
\label{sec:intro}

The Large Ion Collider Experiment (ALICE) \cite{1} at CERN is one of the four main experiments at the Large Hadron Collider (LHC). It studies proton-proton and heavy-ion collisions (such as Pb-Pb) at LHC energies, in order to investigate the Quark Gluon Plasma (QGP) \cite{2}. 
The yields of heavy flavours, \textit{\textit{i.e.}} hadrons containing a quark charm (\textit{c}) or beauty (\textit{b}), and quarkonia, \textit{\textit{i.e.}} bound states $c\overline{c}$ and  $b\overline{b}$, are among the main observables used to study the QGP.

ALICE is equipped with a forward Muon Spectrometer \cite{3},  which detects heavy flavours and quarkonia in their muonic decay channel. It covers the pseudorapidity range -4$< \eta <$-2.5 and it is composed of a hadron absorber, a dipole magnet, a five-station tracking system and the Muon Trigger system (MTR), placed downstream of an iron wall. During Run-I (2010-2013) and Run-II (2015-2018), the MTR was used to identify and select single muons or di-muons with a transverse momentum above a configurable threshold. Muon identification is performed by matching tracklets in the MTR with reconstructed tracks in the tracking stations.

In view of the LHC Run-III (2021-onwards), ALICE will undergo a major upgrade of its apparatus. In order to meet the requirements posed by the forthcoming LHC high luminosity runs, the ALICE experiment will run in continuous readout mode (\textit{i.e.} without trigger) and the Muon Trigger will become a Muon IDentifier (MID).
Moreover, some RPCs have integrated a non-negligible charge with respect to their expected lifetime, and aging effects are starting to appear. For this reasons, a new production of RPCs has been launched. 

In this paper, the details of the MTR upgrades are given. 


\begin{figure}[ht] 
    \centering
    \includegraphics[width=.5\linewidth, height=.14\textheight]{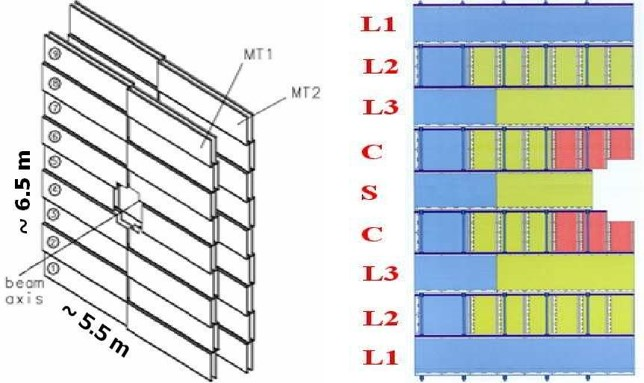} 
  \caption{\label{MTR} Schematic view of the ALICE MTR and detector composition of a half plane. \textit{L1}, \textit{L2} and \textit{L3} indicate the Long RPCs, \textit{C} the Cut ones, and \textit{S} the Short one, while the different colors correspond to a different strip segmentation.}
  \end{figure}

\section{The Muon Trigger}
\subsection{Description of the MTR system}
The ALICE Muon TRigger system (figure \ref{MTR}, left) consists of 72 Resistive Plate Chambers, arranged in 2 stations (MT1 and MT2) of 2 planes each. The two stations are located, respectively, at about 16 m and 17 m from the interaction point (IP). Each plane consists of 18 RPC modules, and covers an area of 5.5 $\times$ 6.5 m$^2$, with a 1.2 $\times$ 1.2 m$^2$ central hole for the beam pipe and its shielding. A single module size is 70 $\times$ 270 cm$^2$. 
There are three different kinds of detector, in order to allow the passage of the beam pipe: Long (\textit{L}), Cut (\textit{C}) and Short (\textit{S}), with different shape and segmentation (figure \ref{MTR}, right). 
Each RPC is equipped with orthogonal read-out strips, in order to provide the spatial information along the \textit{x} and \textit{y} direction (on the plane perpendicular to the beam pipe). The total number of strips is about 21000 and, in order to provide an almost flat occupancy throughout all the detector surface and to keep the momentum resolution roughly constant all over the detection planes, the strip pitches (1, 2 and 4 cm wide) and lengths increase with distance from the beam pipe. 
The detector planes are mounted on mechanical frames that can be moved in order to allow the maintenance and the replacement of the installed RPCs.


\subsection{The ALICE RPCs}

ALICE RPCs \cite{4} are single-gap gas detectors with low-resistivity bakelite electrodes ($\rho \simeq$ 3.5 $\times$ 10$^9$ $\Omega$ cm). Both the gas gap and the bakelite electrodes are 2 mm thick. 
The gas gaps have been manufactured by General Tecnica (Monte San Giovanni Campano (FR), Italy) in 2004-2005, the detectors installation in the ALICE cavern ended in 2007, and the RPCs are operating since the beginning of Run-I (2010).  
The chambers have been flushed with the following gas mixture: 89.7$\%$ C$_{2}$H$_{2}$F$_{4}$, 10$\%$ i-C$_{4}$H$_{1}O$, and 0.3$\%$ SF$_{6}$. 
During Run-I and II, the ALICE RPCs worked in the so-called maxi-avalanche mode, which is characterized by higher gain with respect to the avalanche mode, with a signal of the order of few tens of mV. The analog signals picked-up from the strips were discriminated by A DUal Threshold (ADULT) FEE \cite{5}: this electronics does not have any pre-amplification stage, and the detection threshold applied was 7 mV. 
If the analog signal is higher than the threshold, it is converted into a logical LVDS signal and sent to the local trigger electronics, where it is processed. The applied effective high voltage was about about 10400 V at a pressure of 940 hPa and at a temperature of 20$^\circ$C. 
In these conditions, the mean charge per hit in the gas gap was about 100 pC, while the maximum rate capability was about 100 Hz$/$cm$^2$ \cite{6}.


\subsection{Performance of the ALICE RPCs during Run-I and Run-II}

\begin{figure}[ht] 
  \begin{minipage}[h]{0.53\linewidth}
    \includegraphics[width=.87\linewidth,height=.17\textheight]{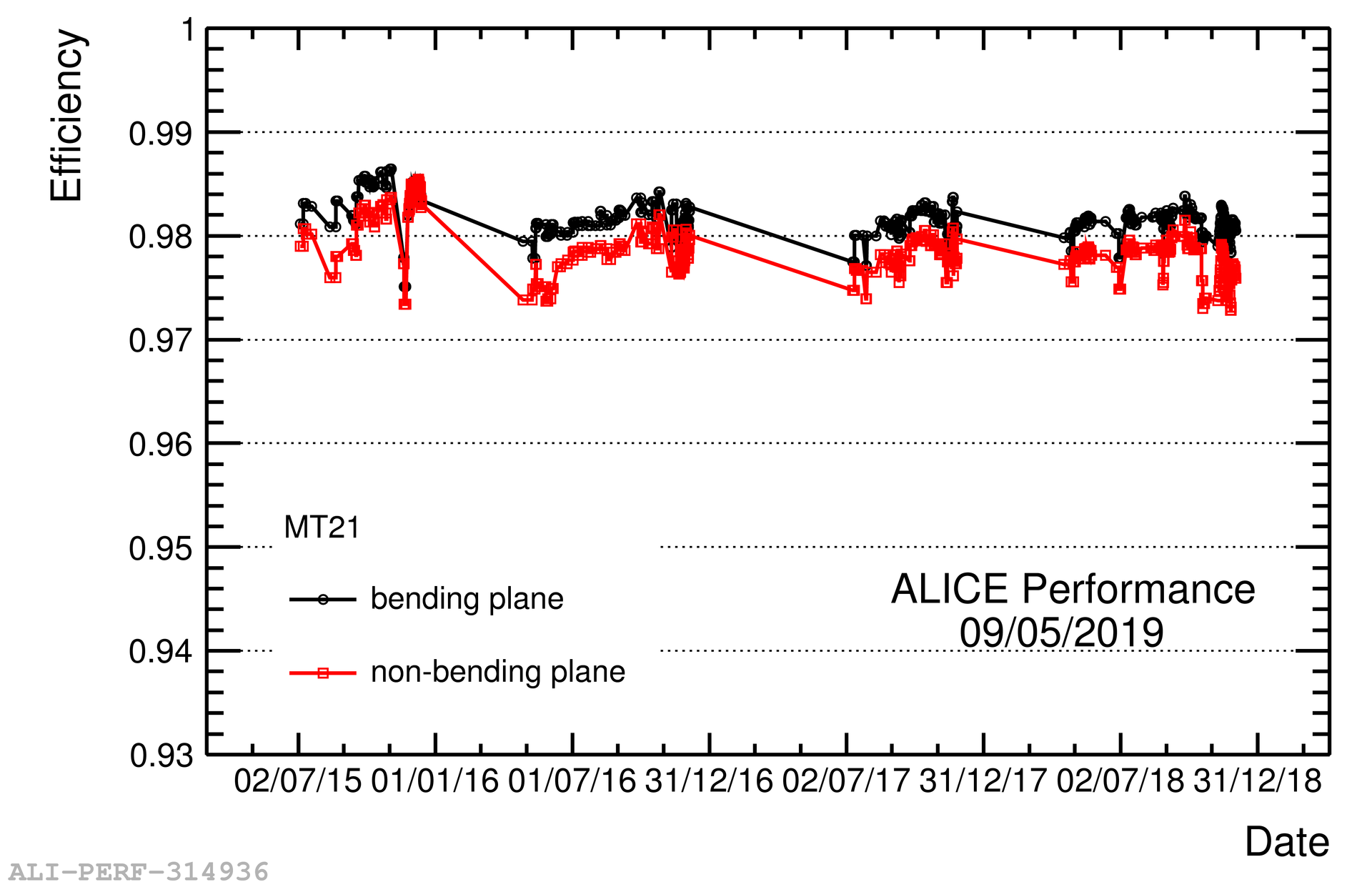} 
  \end{minipage}
  \begin{minipage}[h]{0.58\linewidth}
   \includegraphics[width=.8\linewidth,height=.17\textheight]{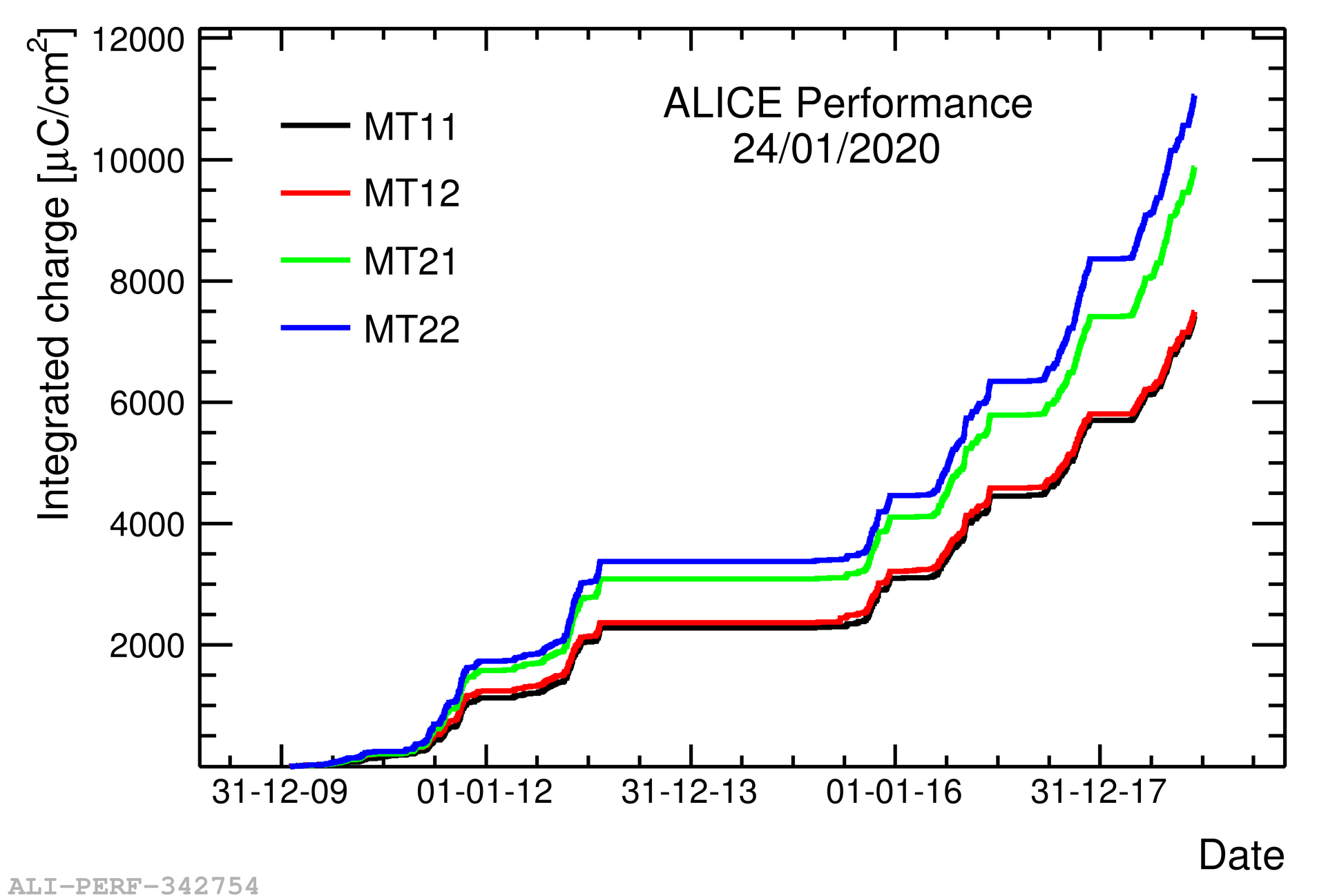} 
    \end{minipage} 
   \caption{Left panel: average efficiency of the MT21 RPC plane during LHC Run-II. Right panel: average integrated charge trend for the four detection RPC planes, during Run-I and Run-II}
   \label{aver_eff}
    \end{figure}

Operational parameters of the ALICE RPCs have been 
monitored throughout all the operation period, and ALICE RPCs  have been stably operated during 
Run-I and II. 
In figure \ref{aver_eff} (left) the trend of the average efficiency for one of the detector planes is shown. The black curve refers to the bending plane (constituted by the strips parallel to the magnetic field), while the red one to the non-bending plane (constituted by the strips perpendicular to the magnetic field). The efficiency 
has been well within the requirements: the average efficiency was $>$ 96$\%$ and stable over time. The small fluctuations are mainly due to the noise in the front-end electronics. 

However, the performance of RPCs can suffer after long-term operation. Ageing effects are roughly proportional to the current drawn during the detector operation period and can be quantified with the time integrated electric charge deposited on the electrodes. 
In figure \ref{aver_eff} (right) the average integrated charge at the end of Run-II for the four detector planes is shown. Some of the detectors have integrated a non-negligible charge with respect to their expected lifetime: the average integrated charge in ALICE RPCs is about 12 mC$/$cm$^2$, while the maximum one is 30 mC$/$cm$^2$. The certified lifetime, measured during R$\&$D studies, is 50 mC$/$cm$^2$ \cite{7}. 


\section{The upgrade: from the Muon TRigger to the Muon IDentifier}

In the forthcoming Run-III, scheduled to start in 2021, some of the ALICE detector will run in continuous readout mode, \textit{i.e.} without trigger, and the MTR play a role of the MID. Moreover, during Run-III, the Pb-Pb collision rate will be increased by a factor 5-10: the maximum counting rate will increase from about 10 Hz$/$cm$^2$ up to 90 Hz$/$cm$^2$. The detectors currently installed in ALICE worked in maxi-avalanche mode with a maximum rate capability of about 100 Hz$/$cm$^2$: this means that, after the LHC upgrade, the maximum counting rate in Pb-Pb collisions will be very close to the maximum rate capability of the RPCs working in maxi-avalanche mode. This increase will also induce an accelerated aging of the detectors, and some of the RPCs installed in ALICE since 2007 would reach the expected lifetime before the conclusion of Run-IV. 
In order to overcome the limitations and reduce the aging, some upgrades 
are necessary.

\subsection{New Front-End Electronics for the RPCs (FEERIC)}

To cope with the increased counting rate, to reduce aging effects and make the detector lifetime comparable to the data taking period of the experiment, RPC detectors will be operated 
at lower gain: in this way the charge released per hit inside the gas gap will be reduced and the rate capability will be increased. This will be possible thanks to new front-end electronics, the FEERIC ASIC (Front-End Electronics Rapid Integrated Circuit) developed at Clermont-Ferrand (France) which, unlike ADULT, will perform an amplification of the analog signal before discrimination \cite{8}. 
Up to now, all the 2384 FEERIC cards (2720 when including the spare cards) have been produced and installed (the installation and commissioning was completed since July 2019). \par
\begin{figure}
    \centering
    \includegraphics[width=.9\textwidth, height=.17\textheight]{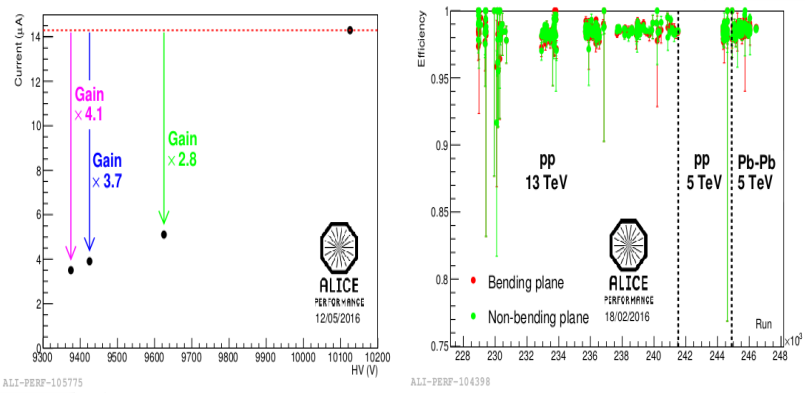}
    \caption{Left panel: current drawn in p-p collisions by the RPC equipped with FEERIC at different HV working point. The gain with respect to the old working point is also shown (the black dot at the right of the plot represents the performance of the same RPC equipped with ADULT cards). Right panel: efficiency of the RPC equipped with FEERIC measured in 2015}
    \label{gaineffFEERIC}
\end{figure}

 
One of the RPCs currently installed in ALICE was already equipped with FEERIC cards from February 2015 until the end of Run-II. This RPCs showed very satisfactory performance and stability \cite{9}. In figure \ref{gaineffFEERIC} (right) the time dependence of the efficiency of the RPC equipped with FEERIC is shown. The efficiency is higher than 97$\%$ during all data taking periods for both the bending and non-bending plane. 
The detector operation with the lower-gain avalanche mode allows us to reduce the working voltage by about 500 V. Furthermore, the reduction in the avalanche charge by a factor of about 5, when compared to the previous operational condition, will dramatically mitigate aging of the gap electrodes by the same factor. 

\subsection{Wireless FEERIC threshold distribution}

Manipulation of digitization thresholds to the front-end electronic is also improved. The ADULT card thresholds were set by means of an analog voltage distribution, and it was possible to set only one threshold value per RPC side. The FEERIC cards will be set via wireless connections, and their values will be set at the single card level, allowing one to tune the threshold according to the local RPC efficiency and to minimize the operating HV.
The ZIGBEE technology has been chosen, which uses the 2.4 GHz radio communication band and is based on Microcontroller Atmel SAMD21; the software is based on Arduino libraries (I2C, SD cards and Xbee). The system is implemented on Xbee cards, designed by the LPC Clermont-Ferrand group. The master cards are connected to the DCS PC via ethernet and the ZIGBEE (wireless) protocol from master to nodes is used. 
There are 26 Xbee cards: the installation of all cards was completed in February 2019 and put into operation.

\subsection{Readout Electronics}

All the readout cards used during Run-I and II will be replaced, in order to switch to the continuous readout mode. There are in total 234 Local cards (16 maximum per VME crates), 16 Regional FPGA-based cards (interfaced with Common Readout Unit via 2 GBTs) and 16 J2 buses between the Local and the Regional cards.  One full crate consists of 16 Local cards, 1 Regional card and 1 J2 bus, and there are 16 similar crates for the full project. All of the cards were designed in SUBATECH, Nantes (France). 
One full readout crate with pre-series cards was validated in spring 2018 and the stress-test at full speed (320 MHz data transfer) was successful. The production of readout cards is completed and all the 16 crates are populated and tested so far. The installation in cavern is ready but postponed due to CERN closure.


\subsection{Test of the new RPCs made with new bakelite}

As explained in Sect. 2.3, some RPCs have integrated a non-negligible charge with respect to their expected lifetime and 
the most aged gas gap will be replaced. In particular, it was decided to replace several RPCs that show a high dark current at the end of Run-II. Although the dark current values are not yet worrisome, it is better to perform the substitution during the long shut down before the beginning of the Run-III, in order to avoid future efficiency losses.
Additional RPCs will be replaced during Run-III, if they show a decrease in the efficiency during data taking. In total, about 25$\%$ of the RPCs currently installed in ALICE will be substituted. 
For this reason, a fresh production of RPC gas gap was launched.
In parallel, an Argon-plasma test is performed, in order to understand and, possibly, revert the increasing trend of dark current \cite{10}. \par 
The new RPCs are made with a new type of bakelite, because the one used for the RPCs currently installed in ALICE is no longer commercially available. 
The new bakelite sheets have, in principle, different surface and bulk properties. 
The RPCs are tested at INFN Technological Laboratory in Torino, with a cosmic rays test station. The tests are made with the streamer mode gas composition (50.5$\%$ Ar, 41.3$\%$ C$_{2}$H$_{2}$F$_{4}$, 7.2$\%$ i-C$_{4}$H$_{10}$, 1$\%$ SF$_{6}$), using the ADULT front-end electronics, with an internal threshold of 80 mV. The tests are performed in this configuration, even if they will operate with the avalanche mixture, because it is necessary to compare the performance of the new detectors with the ones related to the detector currently installed in ALICE, so the experimental conditions must be the same of the tests performed in 2005 \cite{11}.  
The efficiency is measured by means of a cosmic ray tracking station composed of: 3 planes of 9 scintillators each on a moving support, 2 tracking RPCs on the same moving support, and 4 test slots for the RPCs under test, placed between the 2 tracking RPCs. The tracking RPCs and the scintillators form the cosmic trigger system. Only 2 half-RPCs at a time can be tested, because the active area covered by the tracking RPCs is not sufficient to test the entire detector.
The tests carried out are the following: detection of any gas leaks, check of the absence of ohmic leakage currents, the efficiency as a function of HV in cells of 20$\times$20 cm$^{2}$ in order to select the working HV, measurement of the noise map of the detector, and the efficiency map at working point with a granularity of 2$\times$2 cm$^{2}$. \par
\begin{figure}
    \centering
    \includegraphics[width=0.49\textwidth, height=.14\textheight]{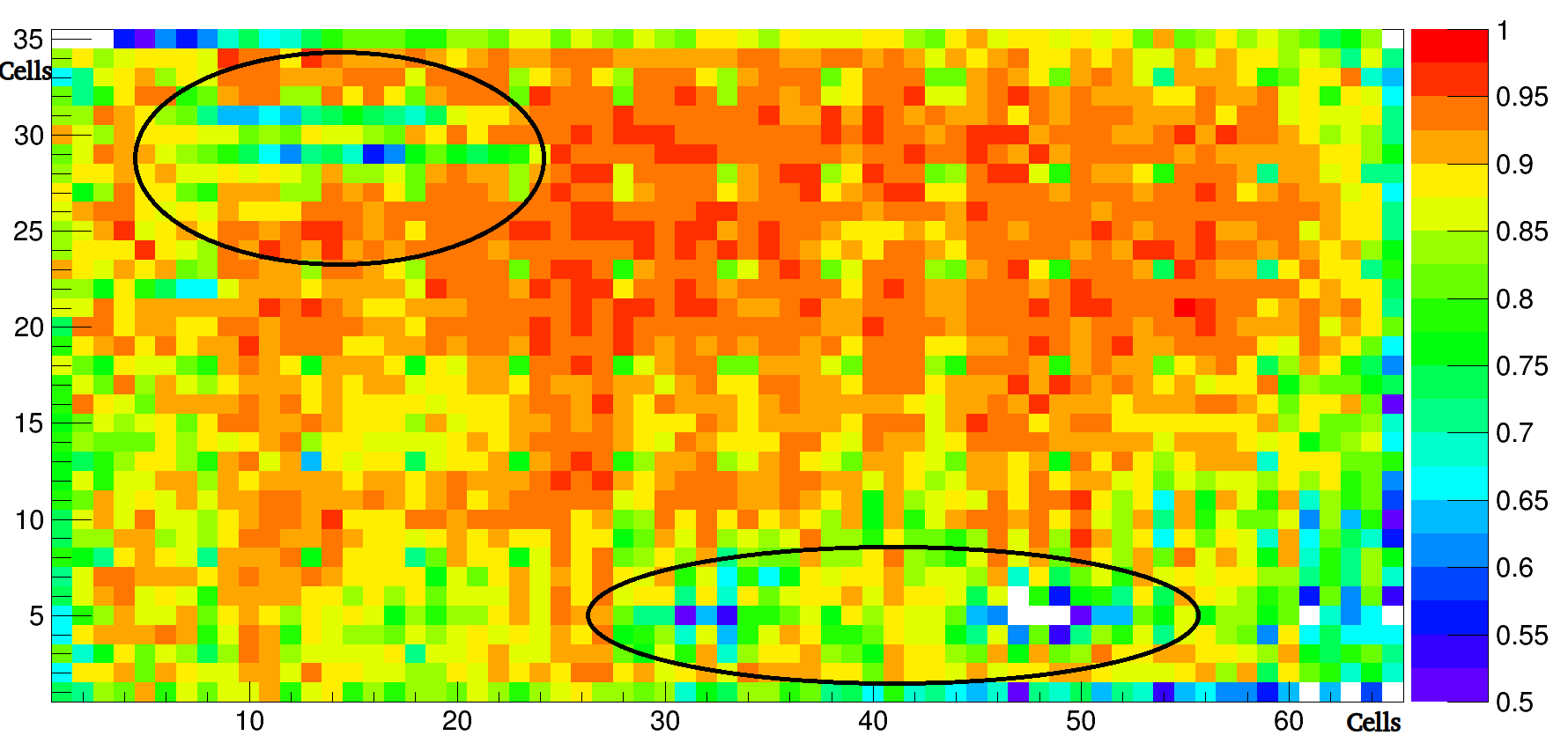}
    \caption{Efficiency map of one RPC half plane at 8700 V, with a granularity of 2$\times$2 cm$^{2}$.}
    \label{highstateff}
\end{figure}

\begin{figure}[ht] 
  \begin{minipage}[h]{0.5\linewidth}
    \centering
    \includegraphics[width=.82\linewidth, height=0.13\textheight]{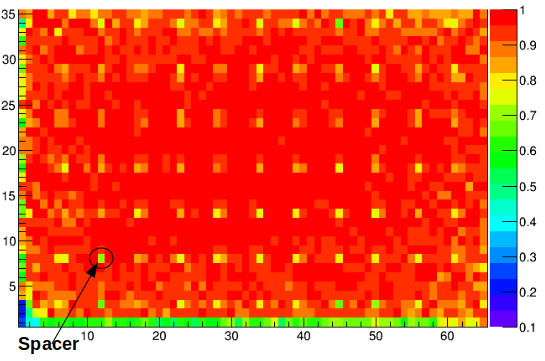} 
  \end{minipage}
  \begin{minipage}[h]{0.5\linewidth}
    \centering
    \includegraphics[width=.82\linewidth, height=0.13\textheight]{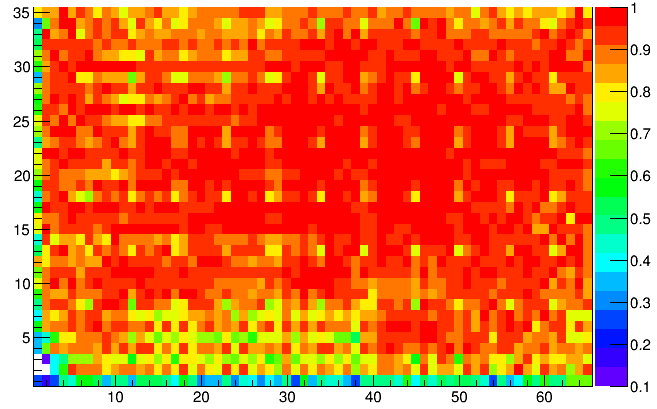} 
  \end{minipage} 
  \begin{minipage}[h]{0.5\linewidth}
    \centering
    \includegraphics[width=.82\linewidth, height=0.13\textheight]{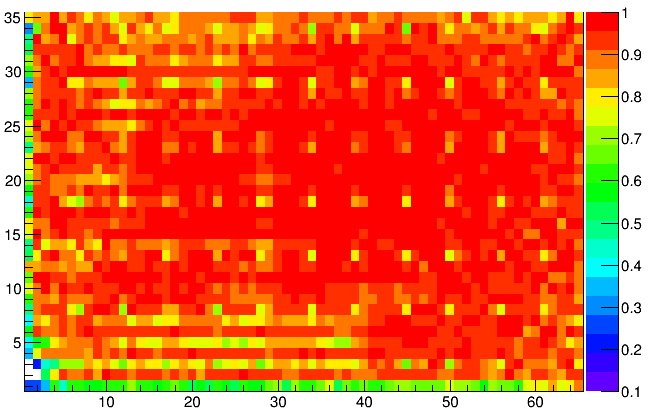} 
    \vspace{4ex}
  \end{minipage}
  \begin{minipage}[h]{0.5\linewidth}
    \centering
    \includegraphics[width=.82\linewidth, height=0.13\textheight]{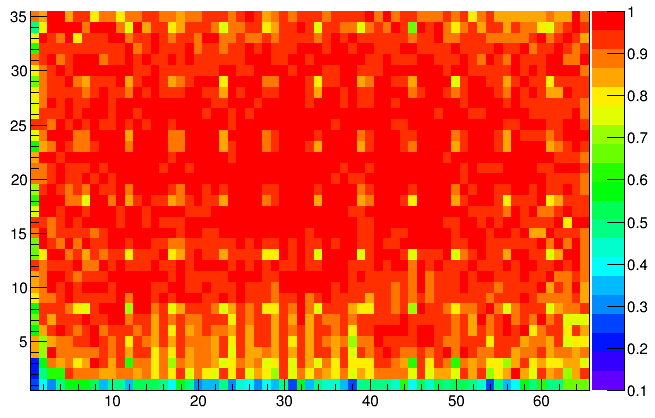} 
    \vspace{4ex}
  \end{minipage} 
  \caption{Upper panel: efficiency map at 8400V, X or Y Strips (left), and efficiency map at 8400V, X and Y Strips (right). Bottom panel: efficiency map at 8400V, X-Strips only (left), and efficiency map at 8400V, Y-Strips only (right).} \label{new}
\end{figure}

 
Up to now, about 30 RPCs have been produced and tested, with a very low acceptance rate. This is mostly due to two factors: the first one is that the new RPCs have an higher working-point HV ($\sim$8700 V), shifted by about 300-400 V, with respect to the RPCs currently installed in ALICE, which have their working point around $\sim$8300 V. The second problem is that the efficiency is non-uniform all over the detector surface (figure \ref{highstateff}). Further investigation showed that the issue was caused by gluing problems in the spacers. For this reason, the gluing procedure is now improved by General Tecnica and a new batch of 3 detectors was delivered in mid-December 2019.
If the tests of these 3 RPCs are successful, the full production will be relaunched.
The test results for one of the newly produced RPC is shown in figure \ref{new}.
The efficiency is estimated by signals occurred in both X and Y strips (top right), in any of the two (top left), or by looking at the X (bottom left) or Y (bottom right) signal separately, in order to distinguish between readout problems and intrinsic gas effects. Only the second configuration gives a uniform efficiency throughout the detector surface.
More tests are ongoing in order to understand if the observation is related to the readout planes or to the gas gap.

\section{Conclusions}
The upgrade of the ALICE Muon Trigger for the forthcoming data taking period at the LHC is ongoing. The new FEERIC FE electronics for the avalanche mode is installed and all tests proved fully satisfactory. 
The production of the readout electronic cards is complete, and all the 16 crates are populated and tested so far. The RPCs that showed the largest dark current values at the end of the Run-II will be replaced. The tests on the new RPC production are ongoing. If the performance will be satisfactory, the production will be started and, in late spring/summer 2020, about 5-10 RPCs in the ALICE cavern will be replaced.

\end{document}